\documentclass[conference]{IEEEtran}
\IEEEoverridecommandlockouts
\usepackage{cite}
\usepackage{amsmath,amssymb,amsfonts}
\usepackage{algorithmic}
\usepackage{graphicx}
\usepackage{textcomp}
\usepackage{xcolor}
\usepackage{hyperref}
\usepackage[export]{adjustbox}
\def\BibTeX{{\rm B\kern-.05em{\sc i\kern-.025em b}\kern-.08em
    T\kern-.1667em\lower.7ex\hbox{E}\kern-.125emX}}
\begin{document}

\title{Bridging the Prototype-Production Gap: A Multi-Agent System for Notebooks Transformation}

\author{\IEEEauthorblockN{1\textsuperscript{st} Hanya Elhashemy}
\IEEEauthorblockA{\textit{Siemens AG} \\
Munich, Germany \\
hanya.elhashemy@siemens.com}
\and
\IEEEauthorblockN{2\textsuperscript{nd} Youssef Lotfy}
\IEEEauthorblockA{\textit{Siemens AG, Technical University of Munich} \\
Munich, Germany \\
yousseflotfy23@gmail.com}
\and
\IEEEauthorblockN{3\textsuperscript{rd} Yongjian Tang}
\IEEEauthorblockA{\textit{Siemens AG} \\
Munich, Germany \\
yongjian.tang@siemens.com}
}

\maketitle

\begin{abstract}
The increasing adoption of Jupyter notebooks in data science and machine learning workflows has created a gap between exploratory code development and production-ready software systems. While notebooks excel at iterative development and visualization, they often lack proper software engineering principles, making their transition to production environments challenging. This paper presents Codelevate\footnote{Source Code: \url{https://github.com/siemens-research/genai-for-sw-architecture}.}, a novel multi-agent system that automatically transforms Jupyter notebooks into well-structured, maintainable Python code repositories. Our system employs three specialized agents — Architect, Developer, and Structure — working in concert through a shared dependency tree to ensure architectural coherence and code quality. Our experimental results validate Codelevate's capability to bridge the prototype-to-production gap through autonomous code transformation, yielding quantifiable improvements in code quality metrics while preserving computational semantics.
\end{abstract}


\section{Introduction} \label{section1}

The past decade has witnessed significant growth in data science and natural language processing (NLP) due to advancements in computational power, the rise of open-source tools, and the increased availability of data. Notebook-based environments such as Jupyter Notebooks and Google Colab have become integral to the data science workflow, offering intuitive platforms for rapid experimentation and visualization. Although these environments support fast prototyping, they can incur technical debt as projects scale beyond initial exploratory analyses.

A core problem arises from the gap between rapid prototyping and sound software engineering practices. Users of notebook environments often work under time constraints, focusing on achieving quick results, which can lead to a neglect of essential design principles like modularity, separation of concerns, and comprehensive documentation. This oversight poses challenges when attempting to transition from exploratory analysis to scalable and maintainable software development.

Generative AI (GenAI) presents an opportunity to address these challenges. Employing a GenAI-powered multi-agent framework can restructure unmaintainable notebook code into a more coherent and scalable codebase, effectively bridging the divide between exploratory data analysis and robust software architecture.

\section{Related Work}
Previous attempts to address the challenges mentioned in section \ref{section1} have mainly relied on standalone Large Language Models (LLMs) or simple automation tools \cite{s1,s2,s3,s4}. However, these approaches often fail to maintain architectural coherence throughout the codebase or ensure consistent application of software engineering principles. The complexity of this transformation task requires a more sophisticated approach that can understand and preserve both local and global code structures while implementing proper software architecture patterns.

Structured multi-agent frameworks are proposed as an alternative, with the potential to surpass the limitations of standalone LLMs. These frameworks emphasize role specialization and modularity, ensuring that each agent within the system operates with a clear and coordinated purpose. This approach mitigates common issues such as shallow abstraction, poor modularity, and inconsistent outputs that often hinder the effectiveness of generative development systems.

Overall, the discussion underscores the importance of advancing beyond traditional LLM-based tools towards agentic systems. Such systems offer structured methodologies that improve code quality and maintainability, effectively addressing the inherent architectural fragility found within notebook-based environments.

\section{System Implementation}
\begin{figure}
  \includegraphics[width=0.5\textwidth]{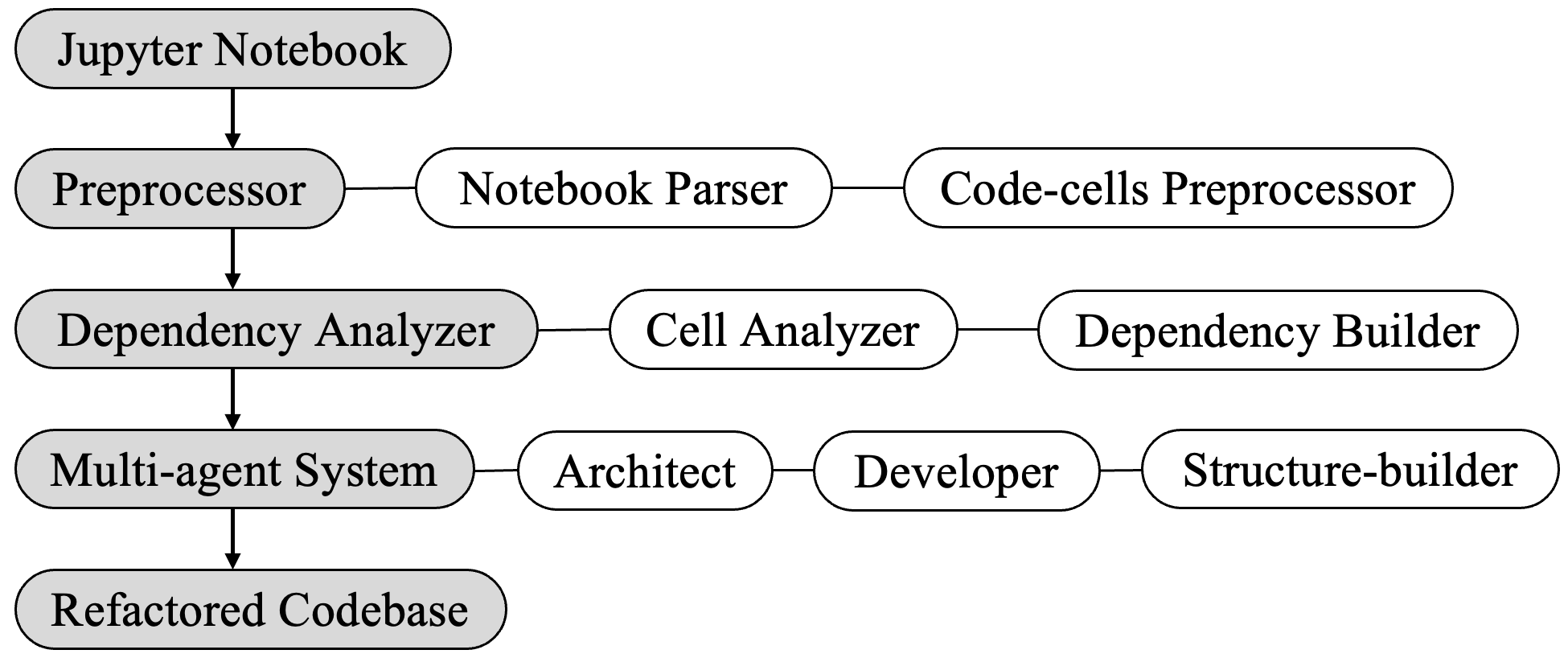}
  \caption{The different stages of the tool's pipeline to transform Jupyter notebooks into modular Python code base}
  \label{fig:Codelevate-overview}
\end{figure}

We developed Codelevate, a Multi-Agent System for Software Architecture (MASA) that automates the transformation of Jupyter notebooks into production-ready Python codebases. This transformation process is structured through a series of well-defined stages that collectively ensure compliance with architectural best practices and increase code maintainability. Figure \ref{fig:Codelevate-overview} shows the different stages of the tool.

In the initial preprocessing stage, the Jupyter Notebook content is parsed and cleaned. The process begins with categorizing cells into markdown, code, and output types using nbformat, a library that facilitates interaction with notebook components. This categorization preserves the semantic integrity of the original notebook format. Subsequently, the code cells undergo tokenization, where they are cleaned of inline comments to strip the content down to its functional logic, thereby preparing it for detailed structural analysis.

Once preprocessing is complete, the focus shifts to dependency analysis. In this stage, a dependency graph is constructed to map out the relationships and interactions among code cells. The abstract syntax tree (AST) module is applied to extract identifiers,
definitions, and usage patterns from each cell. The tree captures the semantic content of the code, including function definitions, variable assignments, and import statements. This graph is crucial for understanding how different sections of code rely on one another, serving as the foundation medium for the code transformation.

The multi-agent component form the heart of the transformation process. Each agent is assigned specialized tasks aimed at enhancing the codebase’s modularity and maintainability. These agents are designed to work in a harmonious sequence, where each stage builds upon the previous one. All agents use the dependency graph as the communication medium. Additionally, the agents, built using LangGraph, leverage OpenAI compatible API, enabling users to choose their preferred LLM with tool-calling capabilities. As an example for this paper we tested the framework using GPT-4o. Each agent reads the graph for context and modifies it based on its assigned tasks, then forwards its modified version to the next agent needed for the transformation. This sequence allows agents to progressively refine the code by applying critical software engineering principles such as DRY (Don’t Repeat Yourself), which reduces redundancy, and SOLID principles, which uphold object-oriented design tenets, like single responsibility and open-closed principles. These practices collectively ensure that the evolving codebase is characterized by robustness and adaptability.

% The multi-agent components is the heart of the transformation process. Each agent is assigned specialized tasks aimed at enhancing the codebase's modularity and maintainability. The agents are designed to work in a harmonious sequence, where each stage of the process builds upon the previous one. The agents all use the dependency graph as the communication medium. Each agent reads the graph for context and modifies it based on its assigned tasks then forwards its modified version to the next agent needed for the transformation. This sequence allows agents to progressively refine the code by applying critical software engineering principles such as DRY (Don't Repeat Yourself), which reduces redundancy, and SOLID principles, which uphold object-oriented design tenets like single responsibility and open-closed principles. These practices collectively ensure that the evolving codebase is characterized by robustness and adaptability.

Finally, in the output generation stage, the transformed code is organized into a structured codebase that features logical file segmentation, thus enhancing both modularity and maintainability. This final product not only maintains the functional equivalency of the original notebook but also surpasses its architectural limitations by providing a clear, maintainable, and efficiently organized code structure.

\subsection{Architect Agent}
\begin{figure}
  \includegraphics[width=0.7\linewidth,height=6cm,trim=-12cm 0 0 0]{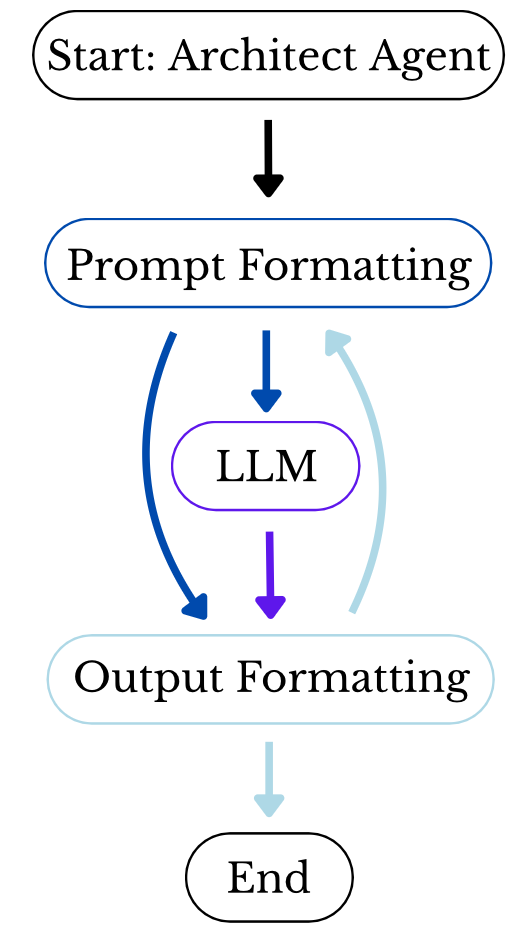}
  \caption{depicts the design of the Architect Agent, generating ADRs for each code cell of a Jupyter notebook and adding the ADRs to the nodes of the code dependency tree.}
  \label{fig:architect}
\end{figure}
The Architecture Agent acts as the primary reasoning component in the MASA framework, designed to analyze parsed and preprocessed Jupyter notebooks represented as dependency trees. Each node within this tree corresponds to a code cell, complete with metadata that includes variable definitions, function usage, and hierarchical parent-child relationships. The agent processes inputs iteratively, focusing on one cell at a time to develop a detailed architectural vision.

The output generated by the Architecture Agent is an Architecture Design Record (ADR) for each cell, documenting specific Architecture Design Decisions (ADDs) aimed at transforming the notebook into a coherent software structure. These decisions encompass modularization strategies, determining whether code should be encapsulated within functions or classes, naming conventions for standardization purposes, strategies for reducing redundancy through the DRY Principle, and the segregation of responsibilities to allocate logic across separate modules or units.

Each ADD is presented with a clear title, a concise transformation description, and references to any parent cell dependencies that inform the architectural decision. The output is structured to ensure it is both parseable and linked directly to the relevant node in the dependency tree, facilitating smooth downstream refactoring. ADRs are integrated back into the corresponding cell nodes to provide guidance for subsequent modifications.

To achieve this architectural transformation, the agent is guided by a system prompt that delineates its responsibilities and architectural objectives. This prompt includes the complete code for the current cell and for all its parent dependency cells, offering the necessary context for developing well-informed ADDs grounded in the original logic and dependencies. The skeleton structure of the dependency tree is also included to model the relationship between current nodes and their dependencies.

The workflow of the Architecture Agent presented in Figure \ref{fig:architect} spans three iterative nodes. The first node, the Prompt Formatting Node, assembles a cell's raw code alongside the codes of its dependencies and structures this information into a well-organized prompt. The second node, the LLM Node, processes the prompt using a LLM of choice to generate an ADR tailored to the cell's architectural context and content. Finally, the Output Formatting Node formats the generated ADR and embeds it into the corresponding cell in the dependency tree.

\subsection{Developer Agent}
\begin{figure}
  \includegraphics[width=0.7\linewidth,height=5.5cm]{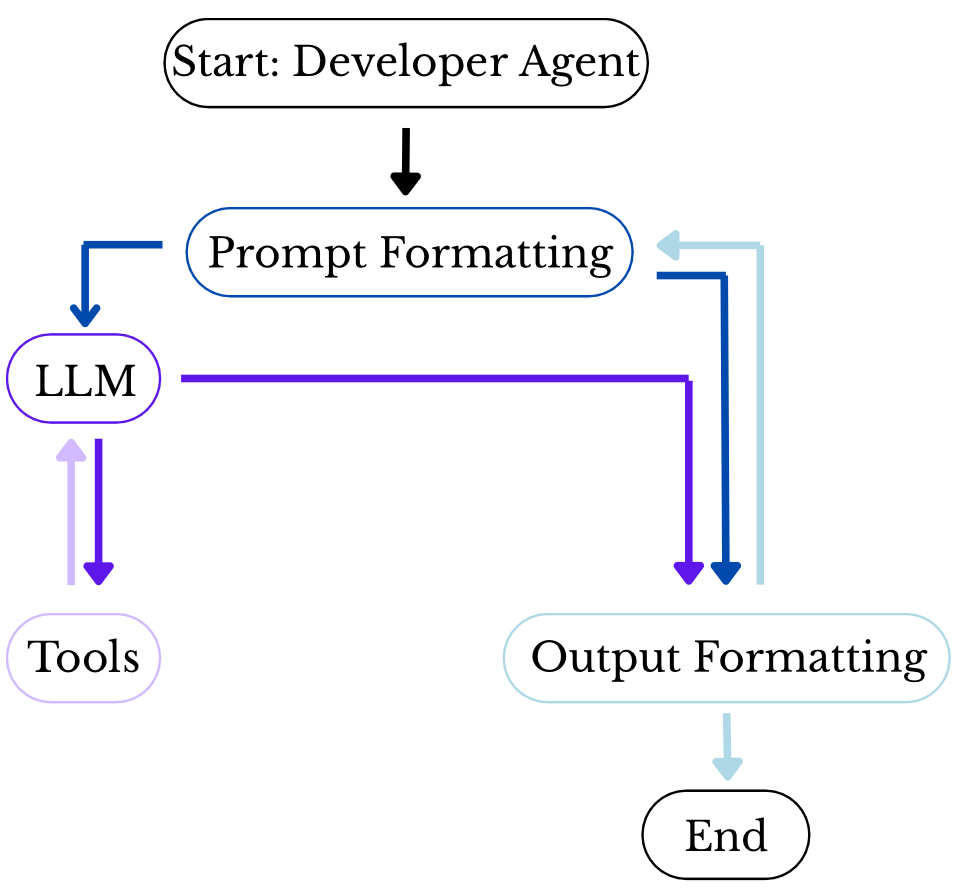}
  \caption{illustrates the design of the Developer Agent, highlighting the various agent nodes and their tool capabilities. The agent iteratively refactors and validates the code before formatting it in the nodes of the dependency tree}
  \label{fig:developer}
\end{figure}
The Developer Agent receives a dependency tree enriched with ADRs and methodically refines each code cell according to the specified architectural guidelines. The agent operates by integrating the raw code of the current cell with the code from its parent cells and the corresponding ADR to construct a detailed prompt. This integration facilitates the alignment of the code structure with the overarching architectural vision while preserving the semantic correctness and functional integrity of the original logic.

As an implementation specialist, the Developer Agent converts architectural intent into concrete code that adheres to design principles, ensuring the refactored code maintains consistency with the original functionalities and dependencies. The output is a refined version of the main cell's code that aligns with the ADR's principles and is prepared for seamless integration into the final codebase.

% To ascertain that the refactored code retains its original functionality and remains executable, the Developer Agent utilizes a Validation() tool. This tool conducts a static code analysis and allows for dynamic testing, identifying and resolving errors iteratively. Discrepancies detected during execution are flagged for correction, ensuring the resultant code is both syntactically and semantically sound.

To ascertain that the refactored code retains its original functionality and remains executable, the Developer Agent utilizes a \texttt{validation} tool. This tool, based on pylint, conducts static code analysis and permits dynamic testing. Pylint provides detailed feedback on coding standards, detects errors, suggests refactoring opportunities, and assigns a quality score to the code. The Developer Agent aims to improve this score iteratively, enhancing the code's overall quality and maintainability. Discrepancies detected during execution are flagged for correction, ensuring the resultant code is both syntactically and semantically sound.

The Developer Agent's workflow presented in Figure \ref{fig:developer} is structured through a four-node pipeline. It begins with the Prompt Formatting Node, which consolidates code from the current cell and its dependencies, coupled with the ADR, to create a complete prompt. This prompt provides the LLM with the full context needed to generate a refactored version of the cell. The LLM Node then processes this prompt to produce the refactored code, iteratively calling a Tool Node to validate the output. The \texttt{validation} tool within this node uses a linter to scrutinize the code for syntax issues and potential code smells. Finally, the Output Formatting Node parses and extracts the validated refactored code, updating the appropriate cell in the dependency tree. This iterative, validation-focused approach ensures that each code cell is transformed into a robust, maintainable component of the overall software architecture.

\subsection{Structure Agent}
\begin{figure}
  \includegraphics[width=0.9\linewidth,height=7cm]{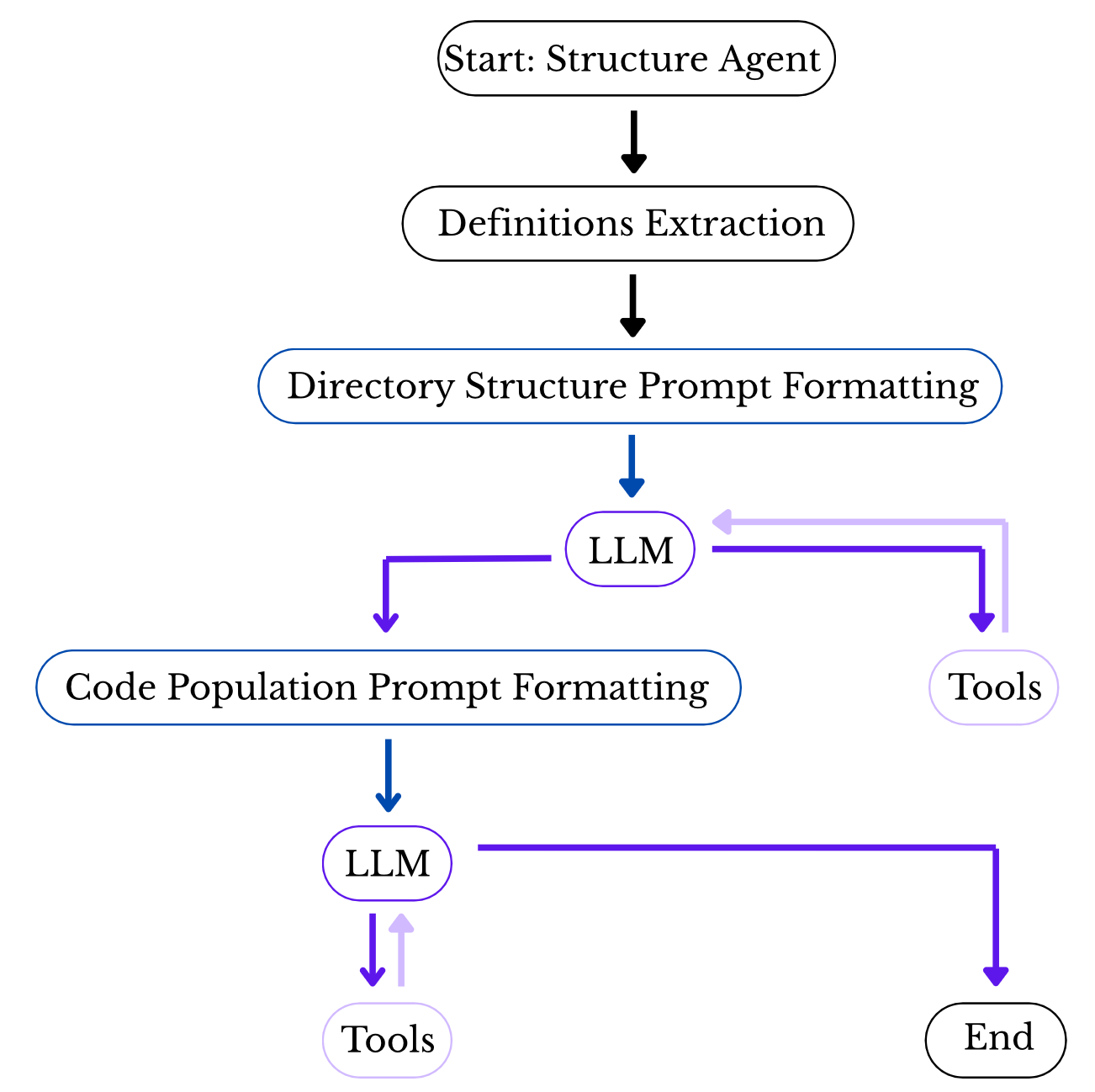}
  \caption{illustrates the design of the Structure Agent, showcasing the various agent nodes along with their tool capabilities to structure and group logically related refactored code into files and modules}
  \label{fig:structure}
\end{figure}
The Structure Agent is the concluding phase in the multi-agent framework, tasked with transforming refactored code into a coherent and well-organized modular structure. It achieves this by utilizing architectural insights and dependency relationships established in preceding stages to enhance maintainability and eliminate redundancy. The agent systematically groups logically related code segments into appropriate modules and files, applying best practices such as the DRY principle to identify and consolidate duplicated functions and classes across cells. This results in a scalable directory layout that significantly improves code readability and promotes reuse.

Using the dependency tree filled by the previous agent with refactored code blocks, the Structure Agent constructs a directory-based file system that mirrors the architectural configuration derived from the dependency tree. Each file within this system is populated with the code corresponding to logically grouped cells. The agent relies on various codebase generation tools like \texttt{write\_file} for writing content, \texttt{read\_file} for reading files, and \texttt{list\_directory} for navigating the directory, alongside \texttt{fetch\_code} to retrieve specific code blocks dynamically.

The Structure Agent workflow presented in Figure \ref{fig:structure} is organized into several nodes, beginning with the Definitions Extraction Node, which traverses each node within the dependency tree to identify structural elements such as function definitions and class declarations. Subsequently, the File Structure Prompt Formatting Node refines these extracted definitions into a prompt format, enhancing the dependency tree skeleton with semantic metadata to guide downstream decisions in file organization and module assignment.

The File Structure LLM Node receives this structured prompt and employs architectural reasoning to devise a directory layout. This layout is physically instantiated using file management tools provided by the Directory Structure Tools Node. The Code Population Prompt Node then translates the dictionary-based file structure into a detailed prompt, mapping file paths to content types, and outlining module responsibilities. This precision guides the Population LLM to populate files within the directory hierarchy.

In its final stages, the Population LLM Node processes enriched prompts to extract refactored code from the dependency tree, ensures logical grouping, consolidates repeated logic under the DRY principle, and writes the code to designated files. Semantic checks are performed to validate execution correctness, maintain architectural coherence, and ensure that each file is devoid of syntax errors and code smells. The Population Tools Node supports this process, empowering the agent with capabilities to write, read, retrieve, and validate code.

The Structure Agent's efforts culminate in a modular, maintainable codebase that reflects coherent architectural transformation. By adhering to best practices in software engineering, it achieves modular design, explicit dependency management, and clear separation of concerns, facilitating improved readability, reusability, and long-term maintainability essential for collaborative development and continuous integration workflows.

\section{Conclusion and Future Work}

Despite the promising outcomes generated by Codelevate, limitations in its current design should be acknowledged. These limitations provide valuable insights for future enhancements and opportunities to expand the system's capabilities. The current system proposes a sequential communication pattern between the Architect, Developer, and Structure agents, which mimics traditional software engineering workflows, but constrains scalability and responsiveness. Existing frameworks, such as HyperAgent\cite{HyperAgent} and Self-Organized Agents \cite{self}, have shown the advantages of asynchronous coordination, suggesting that similar strategies could enhance parallelism and support dynamic task prioritization. Such orchestration approaches could, however, introduce unnecessary overhead, thus a detailed evaluation is necessary before adoption for our use case.

% Additionally, while the system alleviates LLM context window limitations by distributing tasks across specialized agents, more sophisticated transformations requiring long-range consistency could benefit from a shared global memory model
Additionally, while the system distributes tasks across specialized agents, more sophisticated transformations could benefit from a shared global memory model that facilitates long-term consistency between agents. The current reliance on the dependency tree for state transfer lacks dynamic feedback and reinforcement from earlier stages, a challenge that could be addressed by incorporating adaptive memory modules. Handling interactive and graphical output remains open, as the system is primarily geared towards text-based outputs and standard code cells, leaving content such as plots and widgets unsupported.

Furthermore, the system's static refactoring strategy, driven by predefined metadata and prompting strategies, limits its capacity for adaptive optimization across the pipeline. Integrating reinforcement signals from human-in-the-loop scoring could allow the system to iteratively refine its heuristics, leading to more context-sensitive transformations.

Future work involves advancing agent orchestration and embracing parallel execution inspired by task graphs and message queues, as demonstrated by HyperAgent\cite{HyperAgent} or MetaGPT\cite{meta} architectures. Integrating real-time feedback mechanisms could create adaptive learning loops, improving output quality through critic modules or corrective strategies based on prior faults. Extending the framework to manage non-textual and interactive content would bolster its robustness, requiring techniques from UI code generation and reactive frameworks. These insights provide a roadmap for realizing the full potential of leveraging multi-agents for software architecture transformations.


\begin{thebibliography}{00}
\bibitem{HyperAgent} H. N. Phan, T. N. Nguyen, P. X. Nguyen, and N. D. Q. Bui. HyperAgent: Generalist Software Engineering Agents to Solve Coding Tasks at Scale. 2024. arXiv: 2409.16299[cs.SE]. url: https://arxiv.org/abs/2409.16299.

\bibitem{self} Y. Ishibashi and Y. Nishimura. Self-Organized Agents: A LLM Multi-Agent Framework
toward Ultra Large-Scale Code Generation and Optimization. 2024. arXiv: 2404.02183
[cs.SE]. url: https://arxiv.org/abs/2404.02183.

\bibitem{meta} S. Hong, M. Zhuge, J. Chen, X. Zheng, Y. Cheng, C. Zhang, J. Wang, Z. Wang, S. K. S.
Yau, Z. Lin, L. Zhou, C. Ran, L. Xiao, C. Wu, and J. Schmidhuber. MetaGPT: Meta
Programming for A Multi-Agent Collaborative Framework. 2024. arXiv: 2308.00352[cs.AI].
url: https://arxiv.org/abs/2308.00352.

\bibitem{s1} B. Jin, J. Wang, and P. Nie. Suggesting Code Edits in Interactive Machine Learning Notebooks
Using Large Language Models. 2025. arXiv: 2501.09745 [cs.SE]. url: https://arxiv.
org/abs/2501.09745.
\bibitem{s2} A. Shirafuji, Y. Oda, J. Suzuki, M. Morishita, and Y. Watanobe. “Refactoring Programs
Using Large Language Models with Few-Shot Examples”. In: 2023 30th Asia-Pacific
Software Engineering Conference (APSEC). IEEE, Dec. 2023, pp. 151–160. doi: 10.1109/
apsec60848.2023.00025. url: http://dx.doi.org/10.1109/APSEC60848.2023.00025.
\bibitem{s3} B. Liu, Y. Jiang, Y. Zhang, N. Niu, G. Li, and H. Liu. An Empirical Study on the Potential
of LLMs in Automated Software Refactoring. 2024. arXiv: 2411.04444 [cs.SE]. url: https:
//arxiv.org/abs/2411.04444.

\bibitem{s4} T. Sharma. “LLMs for Code: The Potential, Prospects, and Problems”. In: 2024 IEEE 21st
International Conference on Software Architecture Companion (ICSA-C). 2024, pp. 373–374.
doi: 10.1109/ICSA-C63560.2024.00067.

\end{thebibliography}
\end{document}